\documentclass[fleqn,twoside]{article}
\usepackage{espcrc2}
\usepackage{amsmath,epsfig,amssymb}

\usepackage{graphicx}
\usepackage[figuresright]{rotating}

\def\lsi{\raise0.3ex\hbox{$<$\kern-0.75em\raise-1.1ex\hbox{$\sim$}}}
\def\gsi{\raise0.3ex\hbox{$>$\kern-0.75em\raise-1.1ex\hbox{$\sim$}}}

\newcommand{\bq}{\begin{eqnarray}}
\newcommand{\eq}{\end{eqnarray}}
\newcommand{\bqa}{\begin{eqnarray}}
\newcommand{\eqa}{\end{eqnarray}}

\newcommand{\be}{\begin{equation}}
\newcommand{\ee}{\end{equation}}
\newcommand{\bea}{\begin{eqnarray}}
\newcommand{\eea}{\end{eqnarray}}
\newcommand{\half}{\frac{1}{2}}

\title{$N_f=2$ Lattice QCD and Chiral Perturbation Theory}

\author{
 L.~Scorzato\address[HU]{ Institut f\"ur Physik, Humboldt-Universit\"at zu Berlin,
Newtonstr. 15, 12489 Berlin, Germany}\thanks{Speaker}, 
 F.~Farchioni\address[Munster]{Institut f\"ur Theoretische Physik, Universit\"at M\"unster,
Wilhelm-Klemm-Str. 9, 48149 M\"unster, Germany}, 
 P.~Hofmann\addressmark[Munster],
 K.~Jansen\address[DESYNIC]{NIC, Platanenallee 6, 15738 Zeuthen, Germany}, 
 I.~Montvay\address[DESYHH]{DESY, Notkestr. 85, 22607 Hamburg, Germany}, 
 G.~M{\"u}nster\addressmark[Munster], 
 M.~Papinutto\addressmark[DESYNIC]\thanks{Present Address:
INFN Sezione di Roma 3, Via della Vasca Navale 84, I-00146 Roma, Italy}, 
 E.~E.~Scholz\addressmark[DESYHH]\thanks{Present Address: Physics Department,
Brookhaven National Laboratory, Upton, NY 11973 USA}, 
 A.~Shindler\addressmark[DESYNIC], 
 N.~Ukita\addressmark[DESYHH], 
 C.~Urbach\address{Institut f{\"u}r Theoretische Physik, Freie Universit{\"a}t Berlin,
Arnimallee 14, D-14195 Berlin, Germany.}\addressmark[DESYNIC]\thanks{Present Address:
Theoretical Physics Division, Dept. of Mathematical Sciences, University of
Liverpool, Liverpool L69 3BX, UK}, 
 U.~Wenger\addressmark[DESYNIC]\thanks{Present Address:
Institute for Theoretical Physics ETH Z\"urich  CH-8093 Z\"urich, Switzerland}, 
 I.~Wetzorke\addressmark[DESYNIC].
}

\begin{document}

\begin{abstract}
By employing a twisted mass term, 
we compare recent results from lattice calculations of $N_f=2$ 
dynamical Wilson fermions with Wilson  Chiral Perturbation Theory (WChPT). 
The final goal is to determine some combinations of Gasser-Leutwyler Low Energy Constants (LECs). 
A wide set of data with different lattice spacings ($a \sim 0.2-0.12$~fm), different gauge actions 
(Wilson plaquette, DBW2) and different quark masses 
(down to the lowest pion mass allowed by lattice artifacts and including negative quark masses) 
provide a strong check of the applicability of WChPT in this regime and the scaling behaviours 
in the continuum limit.
\vspace{1pc}
\end{abstract}

\maketitle

\section{Introduction}

It was recently recognized ~\cite{Farchioni:2004us,Farchioni:2004fs,Farchioni:2005tu} that 
an important obstacle for the simulation of Lattice QCD with light dynamical quarks comes
from the phase structure of Lattice QCD at small quark masses. Lattice artifacts of $O(a^2)$
may prevent the pion mass from being smaller than a certain minimum value.
Such a scenario had been predicted -- as an alternative to an Aoki phase scenario -- in the context
of Chiral Perturbation Theory for the Wilson fermion action (WChPT) ~\cite{Sharpe:1998xm},
and confirmed numerically in ~\cite{Farchioni:2004us}. These possibilities are not only
a feature of all Wilson-like fermions, but also of staggered fermions 
~\cite{Lee:1999zx,Aubin:2004dm,Hasenfratz:2005ri}. The most disturbing aspect of this problem
is that it may be easily overlooked. In fact very long Monte Carlo histories
have been observed, which eventually turned out to be meta-stable points when jumping to
a higher value of the pion mass ~\cite{Farchioni:2004us}. In the case of pure
Wilson fermions long metastable HMC histories ($\sim 1000$ trajectories) have been 
recently observed (before converging to a single point) even for lattice spacings
as low as $a\approx 0.08$~fm and pion masses $m_{\pi}\approx 300$~MeV ~\cite{Jansen:2005yp}.
Even if dangerous, the problem has a simple solution (although for an additional cost): 
the comparison of simulations at
positive and negative quark masses allows to recognize where meta-stabilities appear.

In this scenario the effect of the gauge action has proved to play an
important role. In particular simulations with the DBW2~\cite{Takaishi:1996xj} 
gauge action showed a considerable decrease of the minimal pion mass~\cite{Farchioni:2004fs}.
In this work we look in detail into this effect. 

The twisted mass fermion approach~\cite{Frezzotti:2000nk} (tmQCD) provides, among other
advantages, the ideal framework for the investigation of the 
(zero-temperature) phase diagram of lattice QCD with Wilson fermions,
see refs.~\cite{Frezzotti:2002iv,Frezzotti:2004pc,Shindler:2005vj}
for reviews on twisted mass fermions in present and past conferences.
On the analytical side WChPT~\cite{Sharpe:1998xm,Rupak:2002sm,Bar:2003mh},
which has been extended to the twisted mass case 
~\cite{Munster:2004am,Scorzato:2004da,Sharpe:2004ps,Sharpe:2004ny,Aoki:2004ta},
offers an efficient tool to interpret the lattice data.

Recently we have collected a large statistics of lattice data at large and moderate
lattice spacing ($a \sim 0.2 - 0.12$~fm), different gauge actions 
(Wilson plaquette and DBW2) and different quark masses 
(down to the lowest pion mass allowed by lattice artifacts and including negative quark masses).
In this work we show explicitly the comparison between these lattice data 
and WChPT. Although for reliable physical predictions definitely smaller lattice 
spacings are needed, also the lattice spacings that we consider here are useful 
as a starting point for the extrapolation to a continuum limit. 
Nevertheless, the results of the present 
work have to be seen as an exploratory study.

By now a good amount of data has been collected also for the 
tree-level Symanzik improved gauge action (tlSym)~\cite{Weisz:1982zw},
which are very promising, but a detailed ChPT analysis is not yet ready.
We will only comment briefly on that at the end.

\section{Lattice simulations on $N_f=2$ QCD}
\subsection{Fermionic and gauge action} 

\label{sec:action}

The lattice action for a doublet of degenerate twisted mass Wilson fermions 
(in the so-called ``twisted basis'') reads 
\begin{eqnarray}
\label{eq:ferm_action}
S_q = && \sum_x \{ 
\left( \overline{\chi}_x [\mu_\kappa + i\gamma_5\tau_3a\mu ]\chi_x \right) + \\
\nonumber
&& - \half\sum_{\mu=\pm 1}^{\pm 4}
\left( \overline{\chi}_{x+\hat{\mu}}U_{x\mu}[r+\gamma_\mu]\chi_x \right)
\} \ ,
\end{eqnarray}
with $\mu_\kappa \equiv am_0 + 4r = 1/(2\kappa)$,
$r$ the Wilson-parameter, set in our simulations to $r=1$, $am_0$
the bare ``untwisted'' quark mass in lattice units ($\kappa$ is the conventional hopping parameter) 
and
$\mu$ the twisted quark mass; we also define $U_{x,-\mu} = U_{x-\hat{\mu},\mu}^\dagger$ and
$\gamma_{-\mu}=-\gamma_\mu$.

The first reason for including $\mu$ is that the fermionic determinant is free from exceptional
configurations when $\mu\neq 0$~\cite{Frezzotti:2000nk}, which helps to perform 
numerical simulations
with light quarks. The second reason is a general $O(a)$ improvement ~\cite{Frezzotti:2003ni}
and the reduced operator mixing~\cite{Pena:2004gb,Frezzotti:2004wz}, when
$\mu_{\kappa}=\mu_{\kappa,{\rm crit}}$.

For the gauge sector we consider the one-parameter family of actions
including planar rectangular $(1\times 2)$ Wilson loops 
($U_{x\mu\nu}^{1\times 2}$):
\begin{eqnarray}
S_g = & \beta &\sum_{x}(c_{0}\sum_{\mu<\nu;\,\mu,\nu=1}^4
\{1-\frac{1}{3}\,{\rm Re\,} U_{x\mu\nu}^{1\times 1}\} + \nonumber \\
+&& c_{1}\sum_{\mu\ne\nu;\,\mu,\nu=1}^4
\{1-\frac{1}{3}\,{\rm Re\,} U_{x\mu\nu}^{1\times 2}\}) \ ,
\label{eq:gauge_action}
\end{eqnarray}
with the normalization condition $c_{0}=1-8c_{1}$.
We considered three cases: i.) Wilson plaquette gauge action, $c_{1}=0$,
ii.) DBW2 gauge action~\cite{Takaishi:1996xj}, $c_{1}=-1.4088$,
iii.) tree-level Symanzik improved gauge action (tlSym)~\cite{Weisz:1982zw}, $c_1 = -1/12$.

The reason why such variations have been explored relies on the experience ~\cite{Aoki:2002vt}
that this may improve the spectrum of the Wilson fermion operator.

\subsection{Analysis of the data in tmQCD}

In this section we review some formulae which are relevant for the analysis of 
lattice data produced with tmQCD~\cite{Farchioni:2004fs}. In fact tmQCD offers new
possibilities for the determinations of the basic QCD parameters and renormalization
factors.

The twist angle $\omega$ defines the chiral rotation relating 
twisted mass QCD to 
ordinary QCD. In the case of the vector and axial currents the rotation reads
(considering only charged currents, $a=1,2$):
\begin{eqnarray}\label{eq:phys_v}
\hat{V}^a_{x\mu} &=& Z_V V^a_{x\mu}\,\cos\omega\, + 
\epsilon_{ab} \, Z_A A^b_{x\mu}\,\sin\omega, \, \qquad
\\\label{eq:phys_a}
\hat{A}^a_{x\mu} &=& Z_A A^a_{x\mu}\,\cos\omega\, +
\epsilon_{ab} \, Z_V V^b_{x\mu}\,\sin\omega, \, \qquad
\end{eqnarray}
where the hatted currents on the l.h.s. denote the chiral 
currents of QCD (physical currents), while the currents on the r.h.s. are the
corresponding bilinears of the quark-field in the twisted ($\chi$-) basis. Note that
the renormalization constants of these bilinears, $Z_V$ and $Z_A$,
are involved. For a given choice of the lattice parameters, 
the twist angle $\omega$ is determined by requiring parity conservation
for matrix elements of the physical 
currents~\cite{Farchioni:2004ma,Farchioni:2004fs}. 
Since unknown renormalization constants 
are involved, {\em two} conditions are required, our choice being:
\begin{eqnarray}\label{eq:cond_v}
\sum_{\vec{x}} \langle \hat{V}^+_{x0}\, P^-_y\rangle&=& 0\ ,
\\
\label{eq:cond_a}
\sum_{\vec{x},i} \langle \hat{A}^+_{xi}\, \hat{V}^-_{yi}\rangle&=& 0\ . 
\end{eqnarray}
The solution of Eqs.~(\ref{eq:cond_v}) and (\ref{eq:cond_a}) with
Eqs.~(\ref{eq:phys_v}) and (\ref{eq:phys_a}) gives a direct 
determination of the twist angle $\omega$ and of the ratio $Z_A/Z_V$ from lattice data, 
see~\cite{Farchioni:2004fs} for details. 
In particular at full twist where $\omega=\pi/2$ the condition reads
$\displaystyle \sum_{\vec{x}}\langle A^+_{x0}\, P^-_y \rangle = 0$.
The full twist situation can be also obtained by requiring the 
vanishing of the PCAC quark mass in the $\chi$ basis.
Both definitions are optimal in the sense of 
\cite{Aoki:2004ta,Sharpe:2004ny,Frezzotti:2005gi}.

The knowledge of the twist angle is necessary for the determination
of physical quantities like the quark mass and the pion decay constant.
The {\em physical} PCAC quark mass $m_q^{PCAC}$ can be obtained from the Ward identity for the 
physical axial-vector current. An interesting possibility is to use 
Eqs.~(\ref{eq:phys_v}, \ref{eq:phys_a}) and parity restoration together with the 
conserved vector current of the $\chi$-fields $\tilde{V}^b_{x\mu}$ for which $Z_V=1$. 
This gives
\be\label{eq:mpcac_phys}
 am_q^{PCAC} = -i
\frac{1}{2\sin\omega}
\frac{\langle\nabla^*_\mu \tilde V^{+}_{x\mu} P^{-}_y\rangle} 
     {\langle P^{+}_x P^{-}_y\rangle}.
\ee
Analogously, for the physical pion decay constant $f_{\pi}$ we use
\be\label{eq:fpi_phys}
af_{\pi}=(am_\pi)^{-1} \langle 0|\hat A^+_0(0)|\pi^+\rangle=
-i\frac{\langle 0|\tilde V^+_0(0)|\pi^+\rangle}{(am_\pi) \sin\omega} .
\ee

Notice that with this definition the lattice determination of $f_{\pi}$ 
has automatically the correct normalization~\cite{Frezzotti:2001du,Jansen:2003ir}.

Finally, the renormalization constant of the vector current $Z_V$ can be determined 
on the basis of the
non-renormalization property of the conserved current $\tilde V_{x\mu}$~\cite{Maiani:1986yj}.
In the case of twisted mass QCD we use
\be\label{eq:zv}
Z_V=\frac{\langle 0|\tilde V_0^+|\pi^+\rangle}{\langle 0|V_0^+|\pi^+\rangle}\ .
\ee

We also observe that in a mass independent renormalization scheme, the renormalization
factors need to be extrapolated to the chiral limit 
($m_q^{PCAC}\rightarrow 0$). For this choice of renormalization
the parity conserving conditions~(\ref{eq:cond_v}),~(\ref{eq:cond_a}), 
hold up to lattice artifacts.

In this work we consider
the lattice data with parameters summarized in Table \ref{table:1}.
The scaling behaviour of the data has already been shown in~\cite{Farchioni:2005ec}.
Here we concentrate on the comparison with ChPT, which is discussed in the next section.

\begin{table}[htb]
\caption{Simulation points.}
\label{table:1}
\begin{tabular}{|l|l|l|l|l|}
\hline
Action & $\beta$ & $a$ [fm] & $a\mu$  & $L^3\times T$   \\ 
\hline
\hline
DBW2   & 0.67    & 0.19     & 0.01    & $12^3\times 24$ \\
\hline
DBW2   & 0.74    & 0.12     & 0.0075  & $16^3\times 32$ \\
\hline
plaq   & 5.1     & 0.20     & 0.013   & $12^3\times 24$ \\
\hline
plaq   & 5.2     & 0.16     & 0.01    & $12^3\times 24$ \\
\hline
plaq   & 5.3     & 0.14     & 0.008   & $16^3\times 32$ \\
\hline
\hline
\end{tabular}\\[2pt]
\end{table}

\section{WChPT versus LQCD}
The extension of WChPT to the case of adding a twisted mass term 
was considered in 
refs.~\cite{Munster:2003ba,Munster:2004am,Scorzato:2004da,Sharpe:2004ny,Sharpe:2004ps,Aoki:2004ta}.
We denote by $L_i$ the usual Gasser-Leutwyler coefficients\cite{Gasser:1984gg}, while
$W$, $\widetilde{W}$ and $W'$ are some combinations of the new LECs associated to $O(m a)$
and $O(a^2)$ lattice artifacts \cite{Rupak:2002sm,Bar:2003mh,Sharpe:2004ny}.

The new LECs depend on the lattice action and also --  in general -- 
on the definition of the mass parameter,  and of the currents. 

If $\kappa_{\rm crit}$ is chosen -- for instance -- from the vanishing of $\cos(\omega)$ 
and if we define ($\mu_R = \mu/Z_P$):
\begin{eqnarray*}
&m_{\chi R} = Z_S^{-1} \left(\frac{1}{2 a \kappa} - \frac{1}{2 a \kappa_{\rm crit}} \right), \; \; 
\rho = 2 W_0 a,\; \; \\
&\chi = 2 B_0 \sqrt{ m_{\chi R}^2 + \mu_R^2}, \\
&\cos{(\omega)} = \frac{m_{\chi R}}{\sqrt{ m_{\chi R}^2 + \mu_R^2}}, \qquad
\end{eqnarray*}
then we have for the pion mass and the PCAC quark mass:

\begin{eqnarray}
m_{\pi^\pm}^2 &=& \chi
+ \frac{1}{32 \pi^2 F_0^2} \chi^2
\ln \frac{\chi}{(4\pi F_0)^2} + \nonumber\\
&&+ \frac{8}{F_0^2}
\{ (4 L_6 + 2 L_8 - 2 L_4 -L_5) \chi^2 + \nonumber\\
&&+ 2 (2 W - \widetilde{W}) \rho\, \chi \cos{(\omega)} + \nonumber\\
&&+ 4 W' \rho^2 \cos{(\omega)}^2  \},\label{mpi-chpt-k} \\
m_q^{PCAC} &=& \frac{Z_P}{2 B_0}
[\chi + \frac{16}{F_0^2}(W\chi\rho\cos{(\omega)}+  \nonumber\\
\label{mpcac-chpt-k}
&&+ 2 W'\rho^2\cos{(\omega)}^2)]
\end{eqnarray}
Similar formulae are available for $f_{\pi}$ and $g_{\pi}$ \cite{Sharpe:2004ny}.
These formulae are valid in the regime where $m_q^{PCAC}/\Lambda_{QCD} \gtrsim a\Lambda_{QCD}$,
where most of our simulated points are located. In the case of a first order phase transition
scenario (as it appears to be the case), the same formulae hold also for smaller masses 
(although some terms can be dropped).
In the regime of very small masses also NLO calculations have been done, leading
to the addition of the $O(a^3)$ terms, \cite{Sharpe:2005rq,Aoki:2005ii}, 
however we will neglect these corrections here.

\begin{figure*}[t]
\includegraphics[scale=0.45]{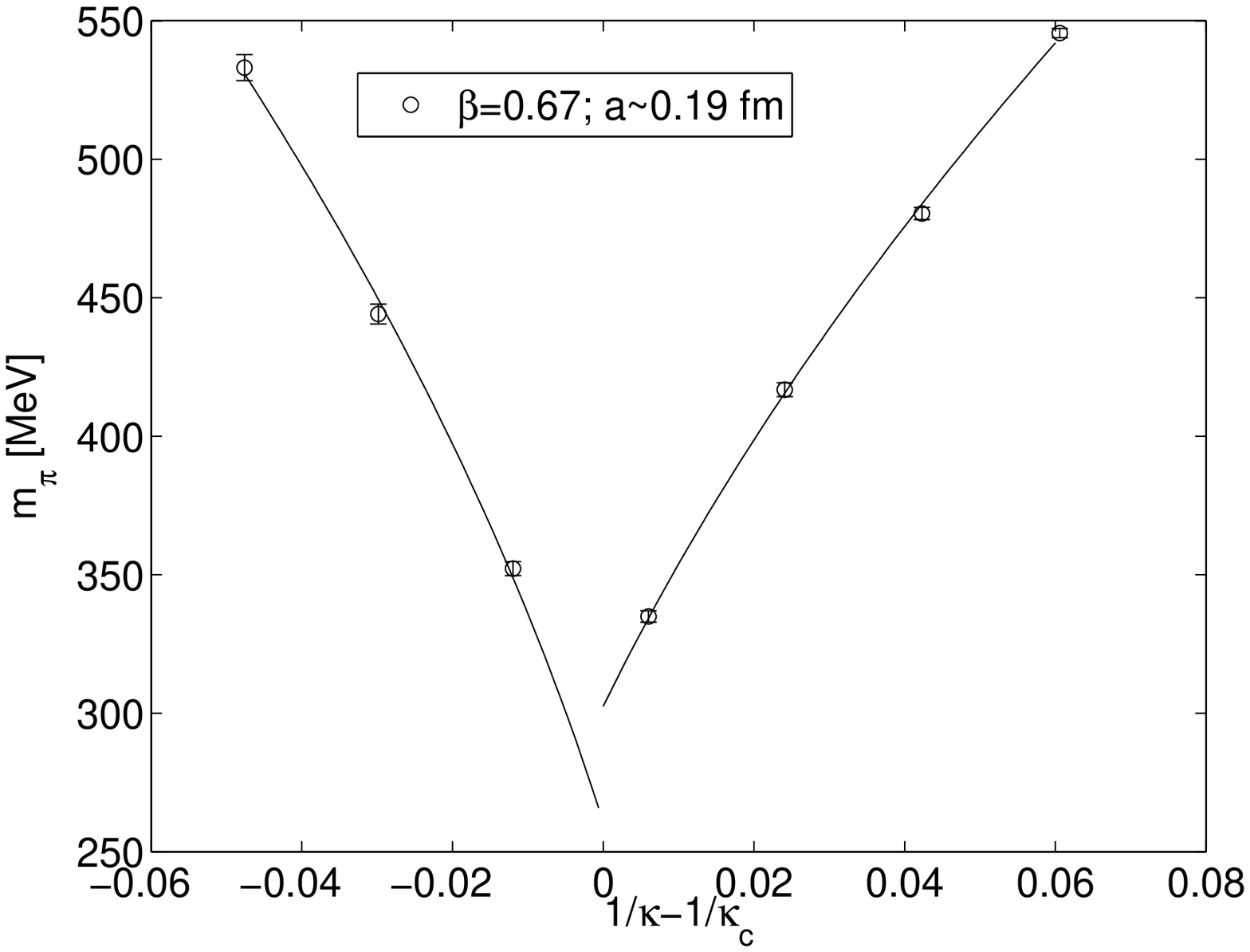}
\includegraphics[scale=0.45]{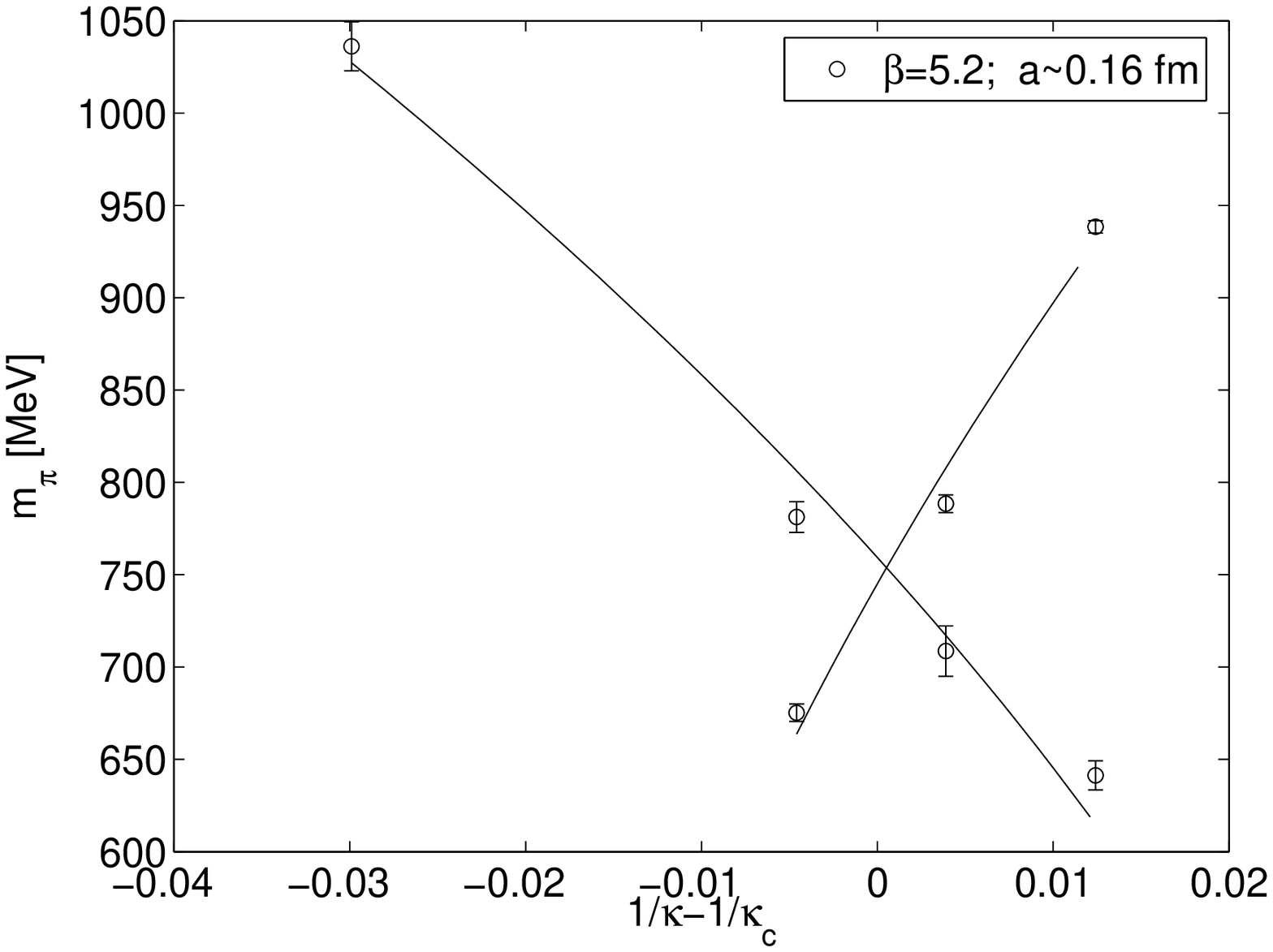}
\caption{
Pion mass for DBW2 (left) and plaquette (right) gauge action at 
$\beta=0.67$ (DBW2), and $\beta=5.2$ (plaquette). For both  V$=12^3\times 24$ 
and $a\mu=0.01$.  Fit of $m_{\pi}^2$ is done with a linear fit.}
\label{fig:mpi}
\end{figure*}

We first show a qualitative comparison of our data with the formulae above.
Figure \ref{fig:mpi} (left) displays the pion mass in the case of DBW2 gauge action 
and a relatively large lattice spacing ($a \sim 0.19$~fm). The data for $m_{\pi}^2$ can be 
reasonably fitted by straight lines (therefore neglecting NLO ChPT terms and the small
twisted mass $\mu$). What cannot be neglected is the presence of a minimal pion mass
of roughly $\sim 300$~MeV. No meta-stabilities have been detected.
If we compare with Figure \ref{fig:mpi} (right) -- whose data 
are produced with plaquette gauge action
at an even smaller lattice spacing ($a \sim 0.16$~fm) -- the striking difference is
the presence of a much higher minimal pion mass. Here many meta-stable points have been
detected, which did not tunnel into stable ones.
Notice that the comparison is done at the same value of $\mu$, and
$a\mu$ is larger for the DBW2 case
(the effect of $\mu$ is however small, at the present value).

\begin{figure*}[th]
\includegraphics[scale=0.45]{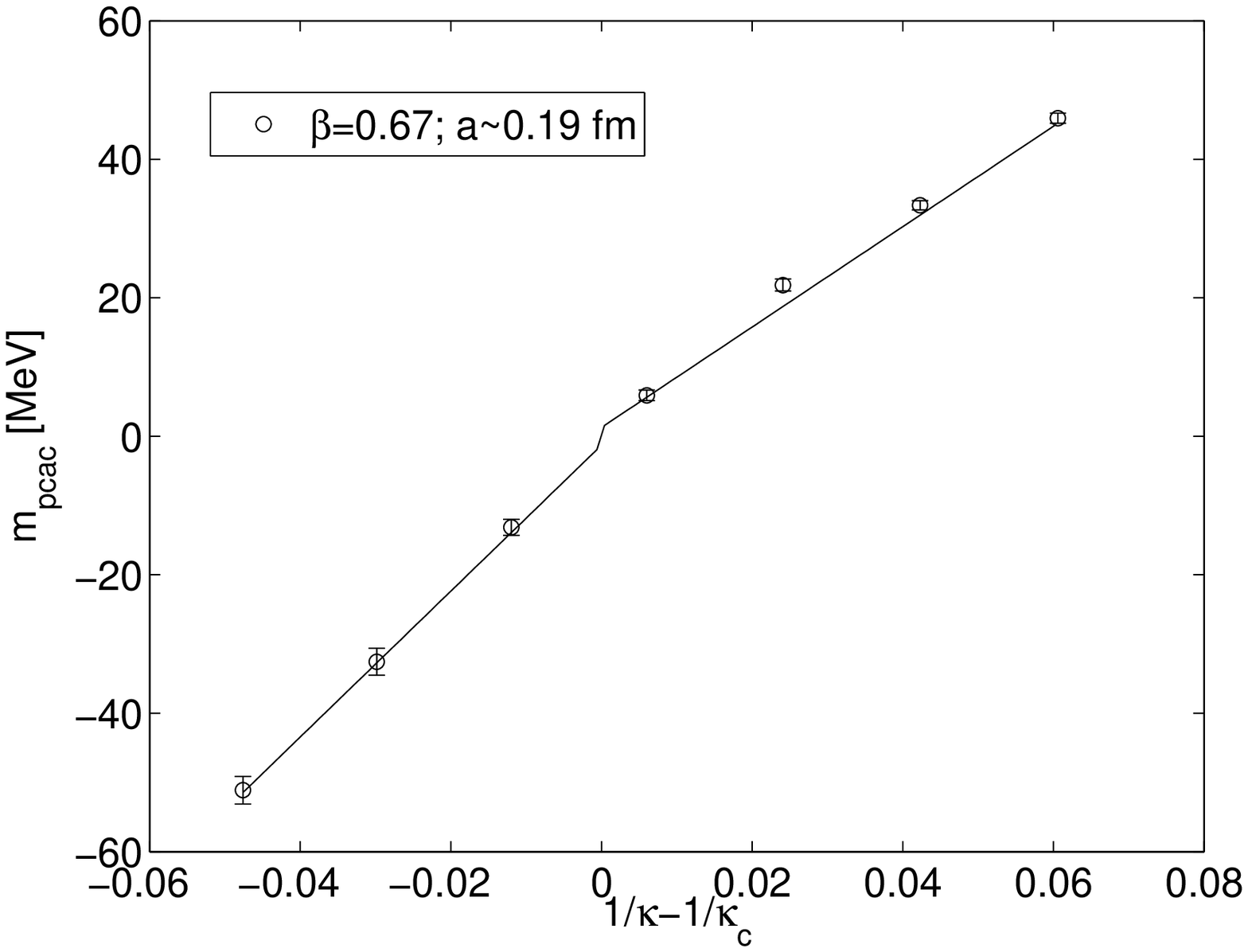}
\includegraphics[scale=0.45]{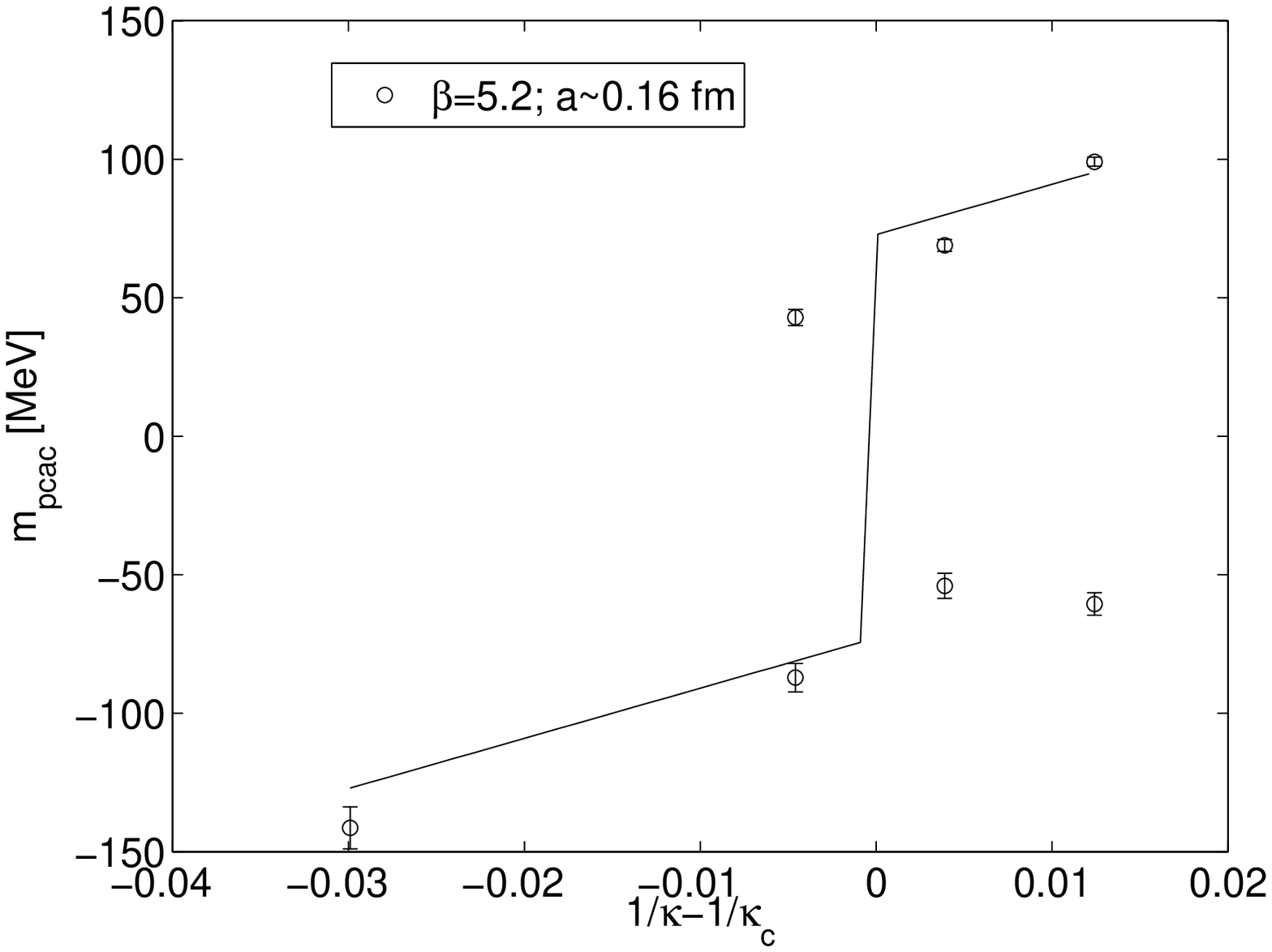}
\caption{
PCAC quark mass for DBW2 (left) and plaquette (right) gauge action at 
$\beta=0.67$ (DBW2), and $\beta=5.2$ (plaquette). For both V$=12^3\times 24$ 
and $a\mu=0.01$.  
Fits are only qualitative, on the basis of Eq. (\ref{mpcac-chpt-k}).}
\label{fig:mpcac}
\end{figure*}

This picture is confirmed by the PCAC quark mass. Eq.~(\ref{mpcac-chpt-k}) shows that
lattice artifacts break the linear relation between $m_q^{PCAC}$ and $\chi$
(they are simply two definitions of the quark mass). In particular the $W$ term
induces a difference in the slope at positive and negative quark masses. The term
$W'$, instead, is responsible for a jump at the origin, which prevents
$|m_q^{PCAC}|$ from being smaller than some $O(a^2)$ lattice artifact.
Both these expected features are visible, but small, in Figure \ref{fig:mpcac} (left): 
two different slopes and a small jump. Also
Figure \ref{fig:mpcac} (right) can be understood in terms of such effects, 
which however completely upset the continuum picture.
The scaling behaviour of these effects is confirmed by the other $\beta$ values 
in Table \ref{table:1}.

The observation that the optimal critical mass can be determined by the vanishing
of $m_{\chi}^{PCAC}$ suggested \cite{Farchioni:2004fs} to express the 
pion mass (and in general all pionic quantities) as function of the PCAC quark mass
instead of $1/\kappa$. The result for $m_{\pi}$ is 
(now $\chi := 2 \frac{B_0}{Z_P} \sqrt{ (\cos(\omega) m_q^{PCAC})^2 + \mu^2}$, 
other definitions are as above):
\begin{eqnarray}
m_{\pi^\pm}^2 &=& \chi
+ \frac{1}{32 \pi^2 F_0^2} \chi^2
\ln \frac{\chi}{(4\pi F_0)^2} + \nonumber \\
&&+ \frac{8}{F_0^2}
\{ (4 L_6 + 2 L_8 - 2 L_4 -L_5) \chi^2 \nonumber \\
&&+ 2 (W - \widetilde{W}) \rho\, \chi \cos{(\omega)} \},
\label{mpi-chpt}
\end{eqnarray}

In this reparametrization the constant $W'$ disappears, 
-- and so does the $O(a^2)$ term -- and the pion 
mass can apparently go to zero when
$m_q^{PCAC}\rightarrow 0$. However, one should keep in mind that not all values 
of $m_q^{PCAC}$ are accessible with stable simulation points.
This parametrization allows to include in the ChPT fit also meta-stable points, where
both $m_{\pi}$ and $m_q^{PCAC}$ are lower
than it would be possible in a stable minimum of the effective potential. 
Since this is an interesting check, we exploit this possibility and we
include also meta-stable points (from \cite{Farchioni:2005tu}) in the  fit.

\begin{figure*}[th]
\includegraphics[scale=0.45]{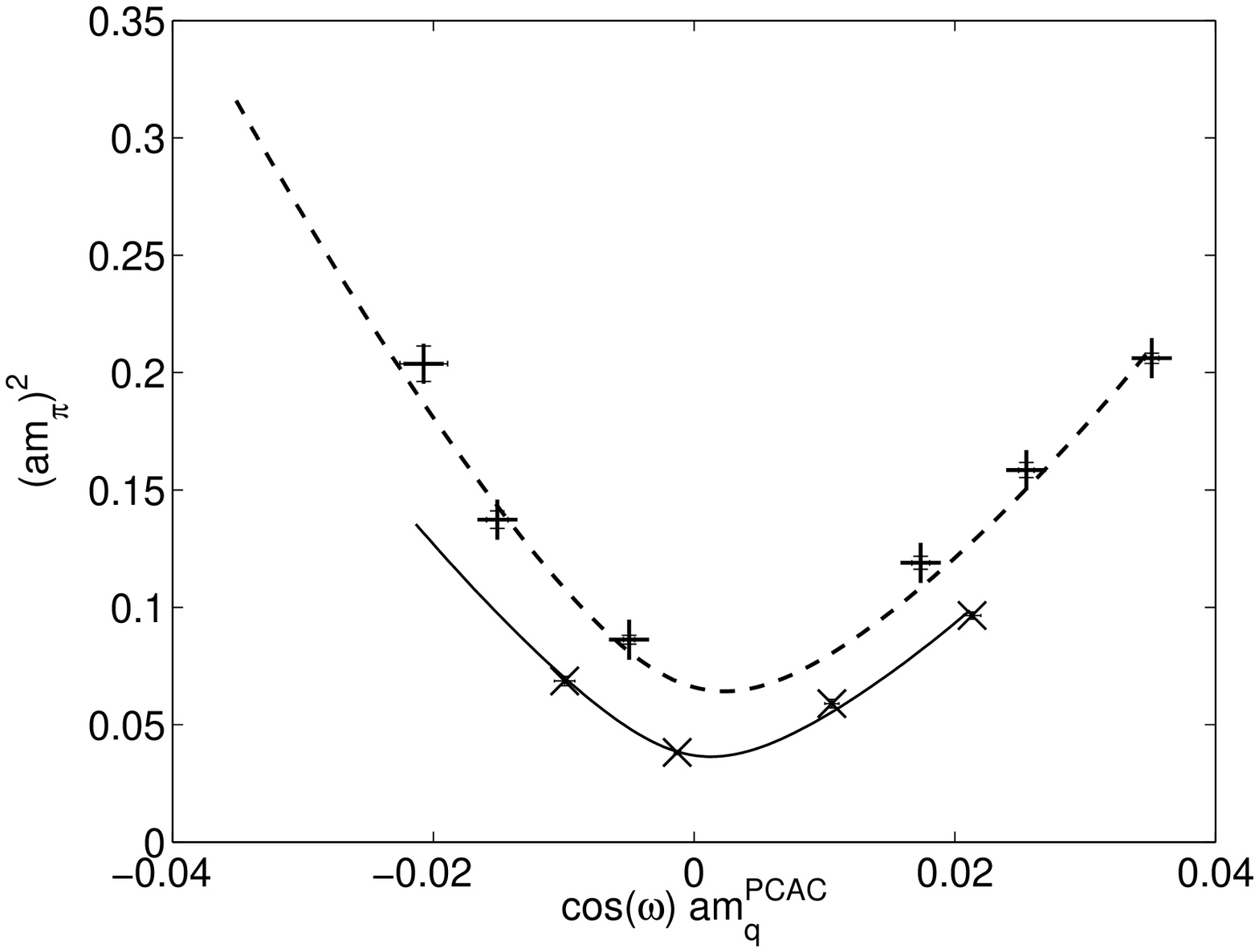}
\includegraphics[scale=0.45]{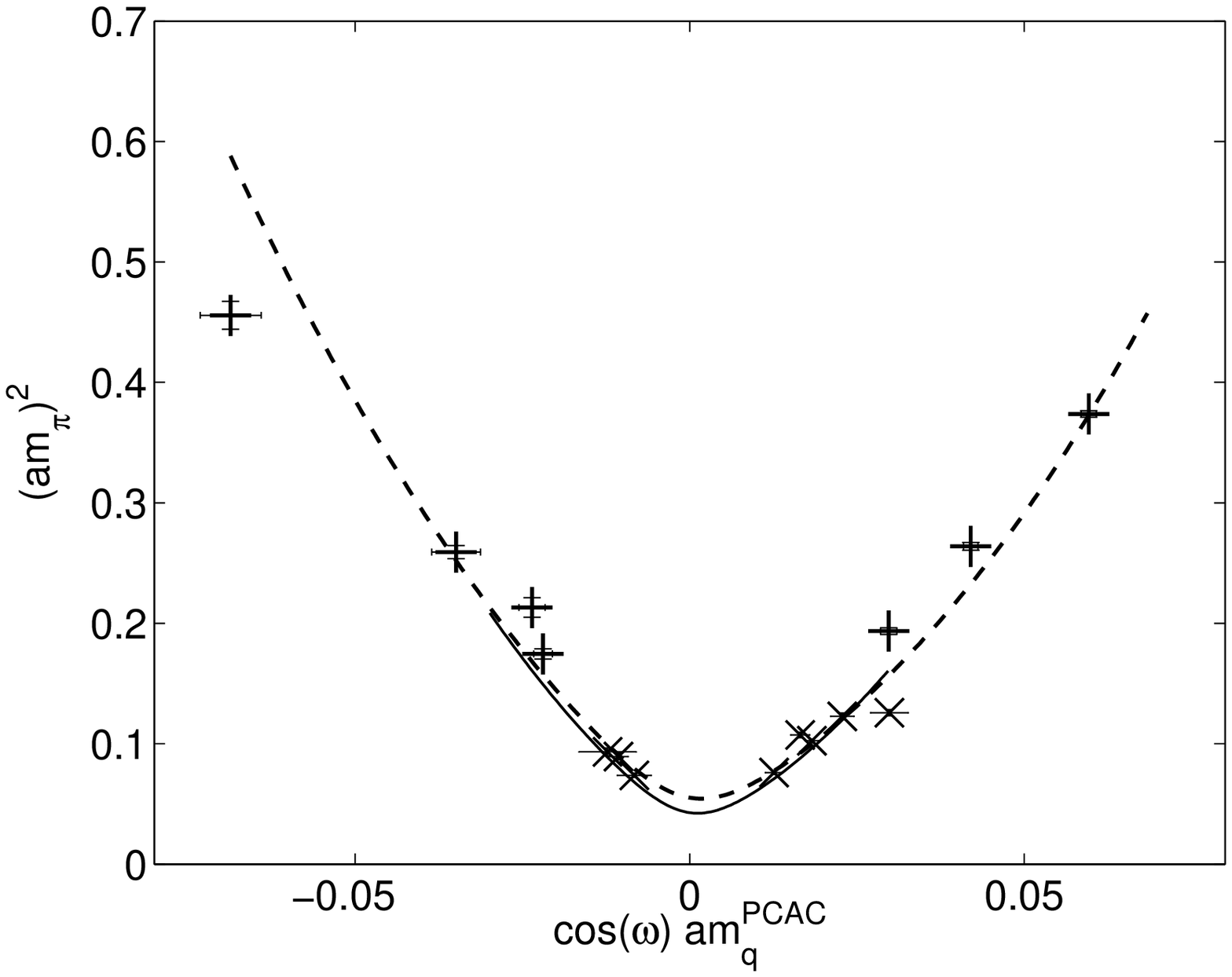}
\caption{Fit of Eq.~(\ref{mpi-chpt}) against the DBW2 data (left) at 
$\beta=0.67$ (dashed line and $+$) and $\beta=0.74$ (full line and $\times$) 
and plaquette data (right) at 
$\beta=5.2$ (dashed line and $+$) and $\beta=5.3$ (full line and $\times$).
Fit are performed in combination with those of $f_{\pi}$ and $g_{\pi}$ 
to provide more constraints, although they are not shown.
}
\label{mpi-pcac}
\end{figure*}

Combined fits of $m_{\pi}$, $f_{\pi}$ and $g_{\pi}$ provide strong constraints.
Results for $m_{\pi}$ are shown in Figure \ref{mpi-pcac}. 
They give an estimate of $F_0\simeq 85$~MeV and values for the LECs in agreement
with previous estimates \cite{Farchioni:2003nf,Farchioni:2004tv}. Results for
the W's appear compatible with zero. Details will be presented elsewhere
\cite{dbw2}.

\section{Conclusions}

We have shown a comparison of unquenched lattice data with WChPT.
We found that WChPT seems to describe the data 
rather well and results for low-energy constants are consistent with previous determinations.
This lets us be confident that a precise determination 
will be possible with reasonable computation cost.

\begin{figure}[htb]
\vspace{-0.0cm}
\begin{center}
\includegraphics[width=1\linewidth]{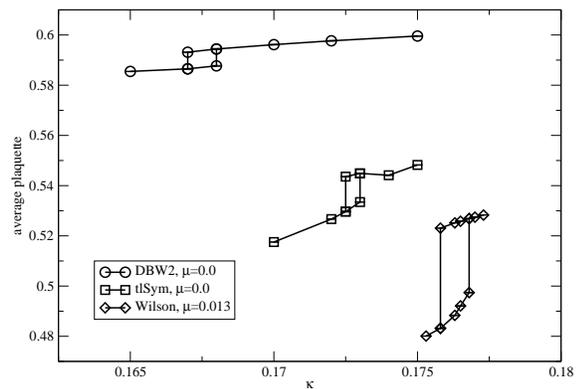}
\end{center}
\vspace{-0.7cm}
\caption{Hysteresis of the average plaquette value as $\kappa$ is moved across
  the critical point, for Wilson, tlSym and DBW2 gauge action at $a\sim 0.17$~fm.
\label{fig:comp_plaq_wilson_dbw2_tlsym}}
\end{figure}

WChPT  lets us investigate the phase structure of lattice QCD. 
There is a striking difference on this aspect between DBW2 and plaquette gauge action.
It is interesting to vary the coupling $c_1$ in Eq.~(\ref{eq:gauge_action}) 
and interpolate between DBW2 and plaquette actions.
It appears that even a small value of $c_1$ can already
have a large impact on the phase structure. This is illustrated in 
Figure \ref{fig:comp_plaq_wilson_dbw2_tlsym}
where we show the average plaquette value as a function of the hopping
parameter $\kappa$ for three different actions, i.e. different values of $c_1$, namely
$c_1=0$ (Wilson), $c_1=-1/12$ (tlSym) and $c_1=-1.4088$ (DBW2).
As one moves $\kappa$ from the negative or positive side across the critical
point, where the PCAC quark mass vanishes, a hysteresis in the average
plaquette value develops whose size and width are indicators of
the strength of the phase transition. We observe that both the width and the
size of the gap in the plaquette value decreases considerably as we switch on
$c_1$ to $c_1=-1/12$ (tlSym action).  Decreasing $c_1$ further down to
$c_1=-1.4088$ (DBW2 action) still seems to reduce the size of the gap, but the
effect is surprisingly small despite the large change in $c_1$. Note that the
results in Figure \ref{fig:comp_plaq_wilson_dbw2_tlsym} are for a lattice
spacing $a\sim 0.17$~fm that is roughly consistent for all three actions.
One should remark that the results for the Wilson plaquette gauge action 
are at non-zero twisted mass $\mu=0.013$, as opposed to the 
tlSym and the DBW2 data in the same plot.
Since the strength of the phase transition is expected to be reduced
as one switches to a non-zero twisted mass, a true comparison at $\mu=0$ would
disfavor the Wilson plaquette gauge action even more. 

Comparison of tlSym data with ChPT will be presented elsewhere \cite{tlSym}.


{\large\bf Acknowledgments}

L.S. wishes to thank the organizers and all the participants of the workshop
for the beautiful and stimulating environment. In particular  
G. Colangelo, M. Golterman, G.C. Rossi and Y. Shamir are acknowledged for 
very helpful discussions.
The computer centers at NIC/DESY Zeuthen, NIC at
Forschungszentrum J{\"u}lich and HLRN provided the necessary technical
help and computer resources. 
This work was supported by the DFG Sonderforschungsbereich/Transregio
SFB/TR9-03.

\bibliography{cyprusproc}

\begin{thebibliography}{10}

\bibitem{Farchioni:2004us}
F.~Farchioni {\em et~al.},
\newblock Eur. Phys. J. {\bf C39}, 421 (2005), [hep-lat/0406039].

\bibitem{Farchioni:2004fs}
F.~Farchioni {\em et~al.},
\newblock Eur. Phys. J. {\bf C42}, 73 (2005), [hep-lat/0410031].

\bibitem{Farchioni:2005tu}
F.~Farchioni {\em et~al.},
\newblock Phys. Lett. {\bf B624}, 324 (2005), [hep-lat/0506025].

\bibitem{Sharpe:1998xm}
S.~R. Sharpe and J.~Singleton, Robert~L.,
\newblock Phys. Rev. {\bf D58}, 074501 (1998), [hep-lat/9804028].

\bibitem{Lee:1999zx}
W.-J. Lee and S.~R. Sharpe,
\newblock Phys. Rev. {\bf D60}, 114503 (1999), [hep-lat/9905023].

\bibitem{Aubin:2004dm}
C.~Aubin and Q.-h. Wang,
\newblock Phys. Rev. {\bf D70}, 114504 (2004), [hep-lat/0410020].

\bibitem{Hasenfratz:2005ri}
A.~Hasenfratz,
\newblock hep-lat/0511021.

\bibitem{Jansen:2005yp}
K.~Jansen, A.~Shindler, C.~Urbach and U.~Wenger,
\newblock hep-lat/0510064.

\bibitem{Takaishi:1996xj}
T.~Takaishi,
\newblock Phys. Rev. {\bf D54}, 1050 (1996).

\bibitem{Frezzotti:2000nk}
Alpha, R.~Frezzotti, P.~A. Grassi, S.~Sint and P.~Weisz,
\newblock JHEP {\bf 08}, 058 (2001), [hep-lat/0101001].

\bibitem{Frezzotti:2002iv}
R.~Frezzotti,
\newblock Nucl. Phys. Proc. Suppl. {\bf 119}, 140 (2003), [hep-lat/0210007].

\bibitem{Frezzotti:2004pc}
R.~Frezzotti,
\newblock Nucl. Phys. Proc. Suppl. {\bf 140}, 134 (2005), [hep-lat/0409138].

\bibitem{Shindler:2005vj}
A.~Shindler,
\newblock hep-lat/0511002.

\bibitem{Rupak:2002sm}
G.~Rupak and N.~Shoresh,
\newblock Phys. Rev. {\bf D66}, 054503 (2002), [hep-lat/0201019].

\bibitem{Bar:2003mh}
O.~B{\"a}r, G.~Rupak and N.~Shoresh,
\newblock Phys. Rev. {\bf D70}, 034508 (2004), [hep-lat/0306021].

\bibitem{Munster:2004am}
G.~M{\"u}nster,
\newblock JHEP {\bf 09}, 035 (2004), [hep-lat/0407006].

\bibitem{Scorzato:2004da}
L.~Scorzato,
\newblock Eur. Phys. J. {\bf C37}, 445 (2004), [hep-lat/0407023].

\bibitem{Sharpe:2004ps}
S.~R. Sharpe and J.~M.~S. Wu,
\newblock Phys. Rev. {\bf D70}, 094029 (2004), [hep-lat/0407025].

\bibitem{Sharpe:2004ny}
S.~R. Sharpe and J.~M.~S. Wu,
\newblock Phys. Rev. {\bf D71}, 074501 (2005), [hep-lat/0411021].

\bibitem{Aoki:2004ta}
S.~Aoki and O.~B{\"a}r,
\newblock Phys. Rev. {\bf D70}, 116011 (2004), [hep-lat/0409006].

\bibitem{Weisz:1982zw}
P.~Weisz,
\newblock Nucl. Phys. {\bf B212}, 1 (1983).

\bibitem{Frezzotti:2003ni}
R.~Frezzotti and G.~C. Rossi,
\newblock JHEP {\bf 08}, 007 (2004), [hep-lat/0306014].

\bibitem{Pena:2004gb}
C.~Pena, S.~Sint and A.~Vladikas,
\newblock JHEP {\bf 09}, 069 (2004), [hep-lat/0405028].

\bibitem{Frezzotti:2004wz}
R.~Frezzotti and G.~C. Rossi,
\newblock JHEP {\bf 10}, 070 (2004), [hep-lat/0407002].

\bibitem{Aoki:2002vt}
Y.~Aoki {\em et~al.},
\newblock Phys. Rev. {\bf D69}, 074504 (2004), [hep-lat/0211023].

\bibitem{Farchioni:2004ma}
F.~Farchioni {\em et~al.},
\newblock Nucl. Phys. Proc. Suppl. {\bf 140}, 240 (2005), [hep-lat/0409098].

\bibitem{Frezzotti:2005gi}
R.~Frezzotti, G.~Martinelli, M.~Papinutto and G.~C. Rossi,
\newblock hep-lat/0503034.

\bibitem{Frezzotti:2001du}
R.~Frezzotti and S.~Sint,
\newblock Nucl. Phys. Proc. Suppl. {\bf 106}, 814 (2002), [hep-lat/0110140].

\bibitem{Jansen:2003ir}
XLF, K.~Jansen, A.~Shindler, C.~Urbach and I.~Wetzorke,
\newblock Phys. Lett. {\bf B586}, 432 (2004), [hep-lat/0312013].

\bibitem{Maiani:1986yj}
L.~Maiani and G.~Martinelli,
\newblock Phys. Lett. {\bf B178}, 265 (1986).

\bibitem{Farchioni:2005ec}
F.~Farchioni {\em et~al.},
\newblock hep-lat/0509131.

\bibitem{Munster:2003ba}
G.~M{\"u}nster and C.~Schmidt,
\newblock Europhys. Lett. {\bf 66}, 652 (2004), [hep-lat/0311032].

\bibitem{Gasser:1984gg}
J.~Gasser and H.~Leutwyler,
\newblock Nucl. Phys. {\bf B250}, 465 (1985).

\bibitem{Sharpe:2005rq}
S.~R. Sharpe,
\newblock Phys. Rev. {\bf D72}, 074510 (2005), [hep-lat/0509009].

\bibitem{Aoki:2005ii}
S.~Aoki and O.~B{\"a}r,
\newblock hep-lat/0509002.

\bibitem{Farchioni:2003nf}
qq+q, F.~Farchioni, I.~Montvay, E.~Scholz and L.~Scorzato,
\newblock Eur. Phys. J. {\bf C31}, 227 (2003), [hep-lat/0307002].

\bibitem{Farchioni:2004tv}
qq+q, F.~Farchioni, I.~Montvay and E.~Scholz,
\newblock Eur. Phys. J. {\bf C37}, 197 (2004), [hep-lat/0403014].

\bibitem{dbw2}
F.~Farchioni {\em et~al.}, in preparation.

\bibitem{tlSym}
F.~Farchioni {\em et~al.}, in preparation.

\end{thebibliography}

\end{document}